\documentclass[final,5p,times,twocolumn]{elsarticle}

\usepackage{amssymb}
\usepackage{lipsum}
\usepackage{graphicx}
\usepackage{float}
\usepackage{lipsum}
\usepackage{CJKutf8}
\usepackage{hyperref}
\hypersetup{
    colorlinks,
    citecolor=black,
    filecolor=black,
    linkcolor=black,
    urlcolor=black
}
\usepackage{cleveref}
\usepackage{soul}
\usepackage[usenames,dvipsnames]{color}

\newcommand{\TSTN}{\mbox{$T_{\rm S}/T_{\rm N}$}}

\def\apgt{\ {\raise-.5ex\hbox{$\buildrel>\over\sim$}}\ }
\def\aplt{\ {\raise-.5ex\hbox{$\buildrel<\over\sim$}}\ }
\def\lteq{\ {\raise-.5ex\hbox{$\buildrel<\over-$}}\ }

\def\aap{\ {A\&A}\ }

\def\aj{\ {AJ}\ }

\def\apj{\ {ApJ}\ }
\def\apjl{\ {ApJL}\ }
\def\apjs{\ {ApJS}\ }

\def\icarus{\ {Icarus}\ }

\def\nat{\ {Nat}\ }

\def\psj{\ {The Planetary Science J.}\ }

\journal{New Astronomy}

\begin{document}

\begin{frontmatter}



\title{A Systematic North-South asymmetry in the Steady-state Climate of rapidly-rotating Oblique Water Worlds}


\author[first]{Yiqi Wu (\begin{CJK*}{UTF8}{gbsn}吴艺琪\end{CJK*})}
\author[first]{Simon Portegies Zwart}
\affiliation[first]{organization={Leiden Observatory,University of Leiden},
            addressline={Einsteinweg 55, 2333 CC}, 
            city={Leiden},
            country={The Netherlands}}
            
\author[second]{H.A. Dijkstra}
\affiliation[second]{organization={Institute for Marine and Atmospheric Research Utrecht,Department of Physics, Utrecht University},
            city={Utrecht},
            country={The Netherlands}}
            
\begin{abstract}
Planetary obliquity (axial tilt) plays an important role in regulating the climate evolution and habitability of water-covered planets.  Despite the suspicion of large obliquities in several exoplanetary systems, this phenomenon remains hard to observe directly. 

We aimed to study the effect of mass, obliquity, and rotation on the steady state climate of water-covered planets. We simulated the climate evolution of such planets with varying obliquities, rotational speed, and mass using a general circulation model (GCM) of intermediate complexity, assuming aqua-planet configurations.

     High obliquity supports an asymmetry between the equilibrium climatological conditions in the northern and southern hemispheres. The polar temperature ratio deviates further from unity with increasing obliquity and rotation rate. Cloud coverage patterns also shift with obliquity, displaying distinct latitudinal bands and increased cloudiness in the warmer hemisphere.

     The climate of habitable-zone aqua-planets turns out to be sensitive to changes in obliquity and rotation rate, but are independent of planet mass.  Our results highlight the importance of considering these factors when assessing the surface conditions of exoplanets.  As a consequence, surface condition     asymmetries in water-world exo-planets can be used to infer the planet's obliquity and rotation rate. 
\end{abstract}




\end{frontmatter}




\section{Introduction}\label{sec:intro}

The discovery of exoplanets between 1\,M$_{\oplus}$ and
10\,M$_{\oplus}$ with stellar irradiation levels between between
$I_{\rm irr} = 1000$ W m$^{-2}$ and $3000$ W m$^{-2}$, initiated the
discussion about habitability of non-rocky planets, broadening the
habitability parameters range. In theory, these planets may have
liquid water-covered surfaces, possibly with a Hydrogen or Nitrogen
atmosphere; these are called ocean worlds
\citep{2008ApJ...673.1160A}. Such habitable planets have an average
temperature at the surface of about 250K to 350K for an albedo $\aplt
0.3$ \citep{benneke2019b, cloutier_18c_2019}.
    
We focus on planets that might possess surface oceans of liquid water
with a mass fraction $\apgt 30$ per cent, such as TOI-1452b
\citep{2021ApJS..254...39G,2023AJ....165..167C} or K2-18b
\citep{k2_discovery}. Such water-rich ocean worlds may be common in
the Milky way Galaxy \citep{2016JGRE..121.1378N}, and beyond
\citep{2008ApJ...673.1160A,Ballesteros2019DivingIE}.

Although ocean worlds tend to reside in a rather restrictive range of
orbital parameters, they remain hard to identify as their surface
tends to hide under a thick cloud cover, making their surface
conditions unobservable.  Top-layer atmospheric conditions of ocean
worlds may exhibit recognizable signatures which can be used to
constrain the parameters of the planet.  We investigate the degree to
which an aqua-planet's rotation speed, obliquity, and atmospheric
conditions are connected.  The former two parameters are hard to
observe directly \citep{snellen2014}, and it is hard to establish a
planet as an ocean world.  The coupling of these parameters, however,
provides a unique and unambiguous proxy for characterizing ocean
worlds.  Using a general circulation model of intermediate complexity,
we establish a relation between lateral temperature asymmetry and the
planet's equilibrium atmospheric conditions, rotation speed, and
obliquity.

\section{Methods}\label{sec:methods}
\subsection{The General Circulation Models of Intermediate Complexity}

General Circulation Models (GCMs) are used to study planetary
atmospheres and climates. Many of these models are geared to Earth's climate, with only limited efforts directed towards modeling exoplanet
atmospheres.

One GCM of intermediate complexity, originally designed for Earth's
atmosphere, is \texttt{SPEEDY} (which stands for \textit{Simplified
  Parametrizations, primitivE-Equation DYnamics}) \citep{Molteni2003,
  OntheNeedofIntermediateComplexityGeneralCirculationModelsASPEEDYExample}.
The code incorporates physically based parameterizations for
large-scale condensation, shallow and deep convection, short-wave and
long-wave radiation, surface fluxes of momentum, heat transport, and
moisture, as well as vertical diffusion \citep{Molteni2003}.  The
simplified parameterizations and relatively low horizontal and
vertical resolution contribute to the computational efficiency of
\texttt{SPEEDY}.

Recently, \cite{Kucharski_ExoSPEEDY}, adapted \texttt{SPEEDY} to model
exoplanet equilibrium atmospheres. The resulting expansion, named
\texttt{ExoSPEEDY}, allows for atmosphere modeling of planets with
parameters and initial conditions slightly different from the Earth.
Although still focusing on Earth-like N$_2$-O$_2$ (Nitrogen-Oxygen)
atmospheres, \texttt{ExoSPEEDY} captures the dynamic interactions
between the atmosphere and the surface ocean.

The adaptation implemented by \cite{Kucharski_ExoSPEEDY} support the
exploration of a wide range of planetary conditions, and allow us to
study exo-planet climates.  We used a version of \texttt{ExoSPEEDY}
with vertical slab ocean model without horizontal diffusion but
including energy transport through the low-resolution grid cells.  The
atmosphere has eight vertical layers from sea level (1013 mbar) to the
base of the stratosphere (30 mbar) with horizontal spectral truncation
T30 ($\approx 3.75^{\circ} \times 3.75^{\circ}$ horizontal
resolution), and T47 \citep[$\approx 2.5^{\circ} \times 2.5^{\circ}$,
][]{Kucharski_ExoSPEEDY}.

\texttt{SPEEDY} adopts observational data from Earth to initialize the
planetary conditions and enforce boundary constraints.  These include
orography, sea-ice fraction, soil wetness, vegetation, sea-surface
temperature, and others.  In \texttt{ExoSPEEDY}, the Earth-specific
profiles can be replaced to prevent bias towards Earth-like climate,
thereby ensuring the integrity and applicability of the model to
exoplanetary environments. In \cref{tab:water_pl_profile}, we provide
an overview of the specific profiles employed in our
simulations. The sea/land mask describes the ratio of ocean coverage
over the entire planet, where 100\% indicates full ocean coverage,
i.e., an aquaplanet. Climatological sea surface temperatures are used
for initialization, and for simplicity we adopt a constant value.

\begin{table}[h]
\centering
\caption{Profiles for initiating a water planet.  Sea/land mask means
  the percentage of land that is covered by ocean, in our case 100\%.
  We adopt a constant climatological sea-surface temperature without
  pre-assigned temperature anomalies. }
\begin{tabular}{@{}lc c}
\hline \hline
Profiles                                & Values [unit]        \\ \hline
Sea/land mask                           & 100$\%$              \\
Climatological sea surface temperatures & 293 [K]           \\
Sea surface temperature anomalies       & Model prediction [K]  \\ \hline
\end{tabular}
\label{tab:water_pl_profile}
\end{table}

\section{Validating the model using an Earth-equivalent}

\begin{table}[h]
\centering
\caption{Selected model parameters. We performed simulations with each
  of the values in obliquity and $\alpha$ for a single value of the
  planet mass (of 8.92\,M$_\oplus$), with all other parameters are
  equivalent to the values for K2-18b in \cref{tab:k2_values}.  An
  additional series of 8 simulations in which the planet mass was
  varied but with an obliquity of $0^\circ$ and $90^\circ$ and a fixed
  $\alpha = 0.0061$ was also performed. For these additional
  simulations, the planet density for the additional mass runs are
  fixed to at 3000 kg m$^{-3}$, and the all other parameters were
  derived accordingly.}
\begin{tabular}{@{}l c}
\hline \hline
Parameter [unit]                    & Range of values \\
\hline
Mass [M$_\oplus$]                    & {1.0, 2.75, 4.5, 6.5, 8.92}\\
Obliquity [$^{\circ}$]               & {0, 23, 35, 45, 55, 90}    \\ 
$\alpha \equiv P_{\rm rot}/P_{\rm orb}$ & {0.083, 0.062, 0.045, 0.03, 0.0061}\\
\hline
\end{tabular}
\label{tab:model_values}
\end{table}

\begin{figure*}[t]
    \centering
    \resizebox{\hsize}{!}
    {\includegraphics[scale = 0.465]{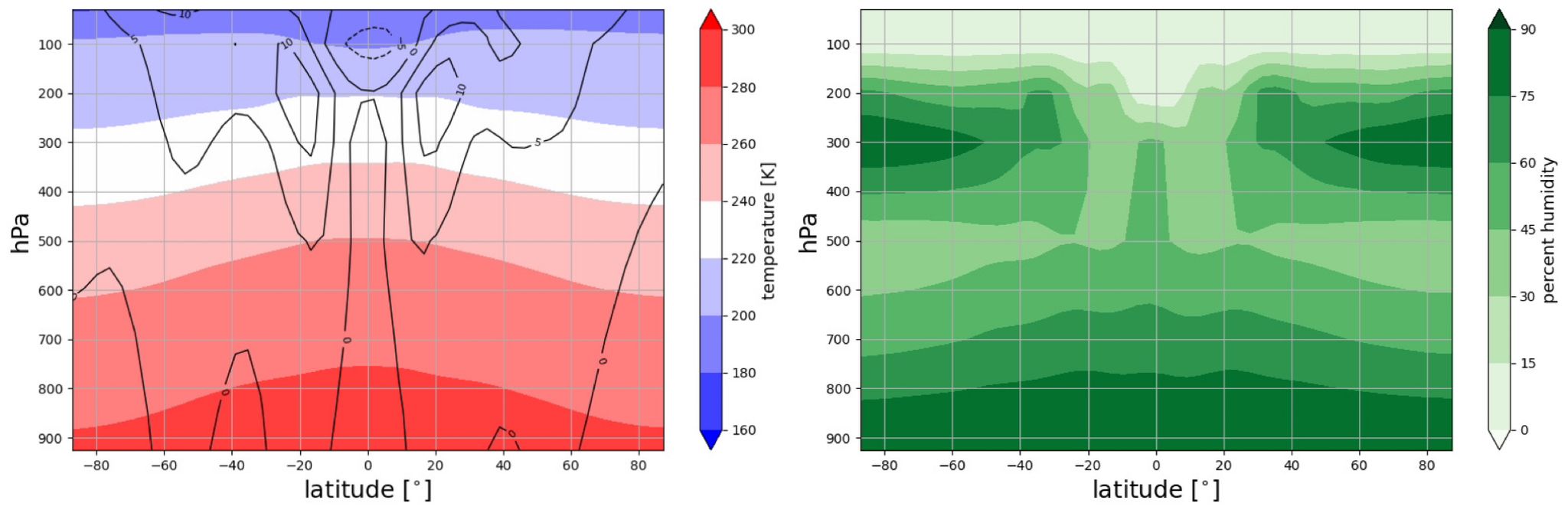}}
    \caption{Zonal-average pressure-latitude for the climate of
      water-Earth with 23.4$^\circ$ obliquity. Latitude 0$^\circ$ is
      at the equator with positive values for the Northern Hemisphere.
      These are the time-average results from the last three years
      of the model simulation.  \textit{Left panel:} Color-map gives
      the average temperature, overplotted with zonal wind velocity
      contour lines (m s$^{-1}$). Dashed lines indicates easterlies
      whereas solid lines represents westerlies. \textit{Right panel:}
      Color-map of the relative humidity.  }
     \label{fig:earth_tw}
\end{figure*}

We validate the numerical setup by performing a detailed comparison to
earlier published results using the Earth as a template.  For this, we
perform a series of calculations with a constant obliquity of
$23.4^\circ$, and an eccentricity $e=0.02$, similar to Earth's. In
\cref{fig:earth_tw}, we show the zonal average pressure-latitude
analysis for the temperature, wind velocity (contour lines in the left
panel), and relative humidity of a water Earth prototype.

The temperature is highest at the equator near the surface and
decreases toward the poles. The stratification of temperature with
pressure indicates a well-defined troposphere around 200 K. The zonal
wind contours reveal distinct regimes. In the equatorial region,
eastward winds (dashed lines) dominate in the upper atmosphere,
indicative of strong equatorial jets. Weaker wind velocities occur
towards the polar region and in the lower atmosphere.

The relative humidity ranges from 0$\%$ to 90$\%$.  The highest
relative humidity is observed near the equator and at lower altitudes,
extending up to the troposphere. This region coincides with the
Intertropical Convergence Zone (ITCZ), where convective activity and
precipitation are intense.  The areas at higher altitudes, close to
the stratosphere, show lower relative humidity, consistent with
subsiding air masses and reduced convective activity.

The relatively symmetrical distribution with equatorial peaks and
decreasing values towards the poles is evident in all of these
observables.  This pattern exists in both temperature and relative
humidity, showcasing pronounced climatic activity in the equatorial
regions, that diminishes toward the polar regions.  Our simulations
give somewhat smaller upper-limits for wind velocity and relative
humidity when compared to \cite{Blackburn2013} and
\cite{earth-climate}, but are consistent in general.  We attribute the
different upper limits to our lower resolution.

We further quantify the uncertainty in our model calculations, using
water-Earth, by performing additional experiments across a range of
eccentricities, obliquities, and semi-major axes. The variations in
orbital elements result in a Solar irradiance ranging from $I_{irr}
\simeq 950$ to $2140$\,W m$^{-2}$, necessary for sustaining liquid
phase H$_2$O on the planetary planet's surface.  These calculations
did not lead to excess evaporation or to the formation of sea ice,
validating our choice of water-covered planets.

\subsection{Initial conditions for water worlds}\label{Sect:WaterEarth}

Having established that the model compared favorably to earlier
water-covered Earth models, we now discuss the more general case of
the effect of planet mass, obliquity, and rotation period on the
climate of water-rich planets.  The motivation for this choice of
parameters stems from the possibility of habitability and the
frequency at which those worlds are found.

\begin{figure*}[t]
    \centering
    \resizebox{\hsize}{!}
    {\includegraphics[scale = 0.465]{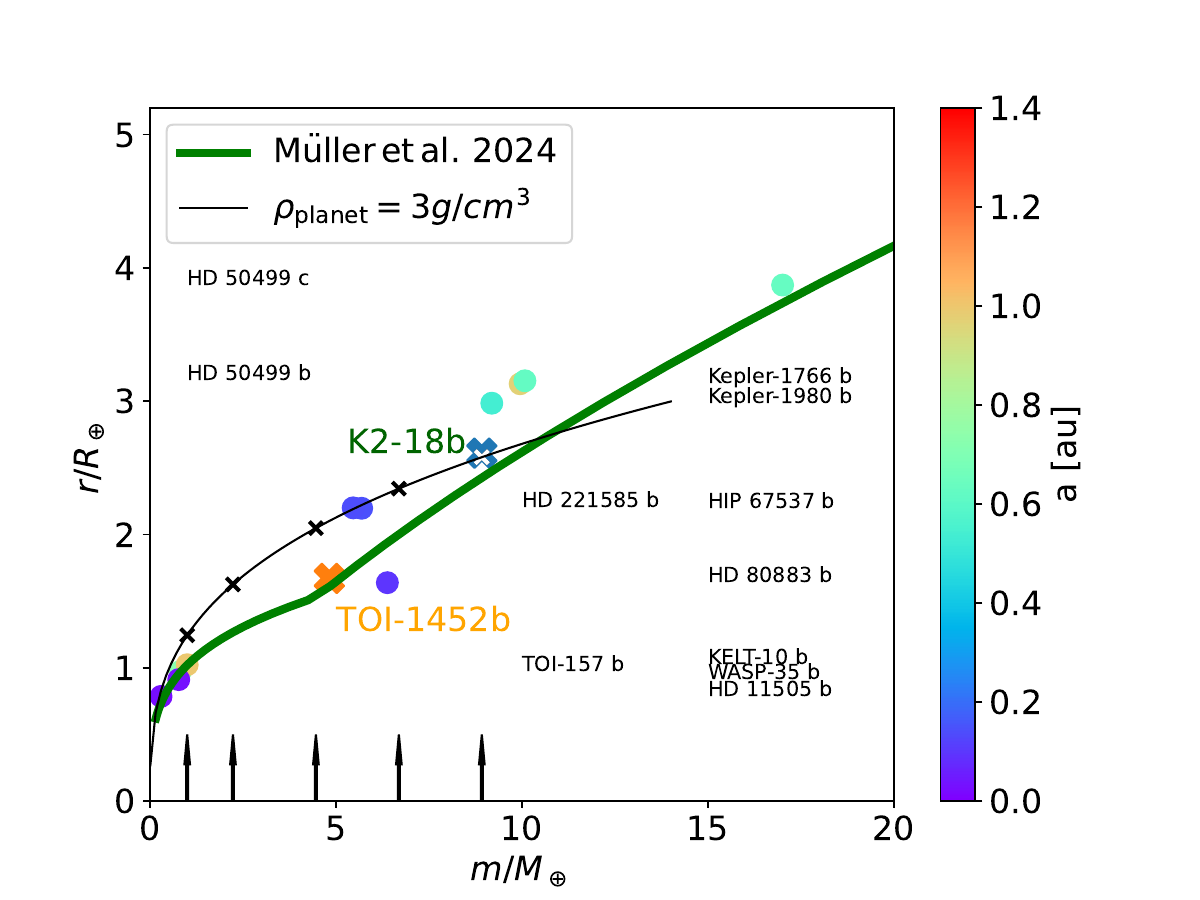}}
    \caption{Mass and radius for a number of potentially habitable
      ocean worlds.  The bullet points show identified ocean worlds:
      listed from bottom to top: HD 11505b, WASP 35, TOI 157, {\bf TOI
        1452b}, KELT 10b, HD 80883b, HD 221585b, HIP 67537b, {\bf
        K2-18b}, Kepler 1980b, HD 50499b, Kepler 1766b, HD 50499c.
      The green curve gives the mass-radius relation for planets with
      a water content of $\apgt 30$\,\% (by mass) from
      \cite{2024A&A...686A.296M}.  The arrows (and black crosses
           [right-most one is white]) give the simulation atmosphere
           models.}
     \label{fig:observed_waterworlds}
\end{figure*}

We simulate water-covered planets with Earth-like atmospheric
compositions and without sea ice.  We assume an Earth-like atmosphere
of 1 bar at sea level, and a N$_2$+O$_2$ atmosphere with 0.04\%
CO$_2$. The latter is used to calculate atmospheric absorption and
transmissivity.  In \cref{fig:observed_waterworlds} we present the
mass-radius relation for a number of observed water-rich planets from
the Open Exoplanet Catalog\footnote{see
\url{https://openexoplanetcatalogue.com/}}.

In this study we focus on TOI-1452b and K2-18b, both of which are
water rich and sufficiently close to their parent star too be warm
engough to be deprived of sea ice.  We list their characteristics in
\cref{tab:k2_values}, and the more specific model free parameters in
\cref{tab:model_values}.  In these calculations we vary the rotational
period of the planet, its mass, and obliquity. 

In \cref{fig:observed_waterworlds} we also present the parameter space
for water-worlds, and identify our selected initial parameters with
the up-pointing arrows.  Although motivated by TOI-1452b and K2-18b,
our selected parameters are general for water-covered planets with an
Earth-composition atmosphere.  In \cref{tab:k2_values}, we list the
parameters for these planets, and in \cref{tab:model_values} we
indicate the variation over which we cover planet mass, obliquity and
rotational period.

\begin{table}[h]
\centering
\caption{Parameters TOI-1452b and K2-18b.}
\begin{tabular}{@{}lc cc}
\hline \hline
Parameter [unit]                    & TOI-1452b & K2-18b \\
\hline
Stellar mass [M$_\odot$]             & 0.249     & 0.495 \\
Semi-major axis [au]                & 0.061     & 0.159  \\
Eccentricity                        & 0.0       & 0.09    \\
Temperature [K]                     & 326       & 300     \\
Radius [R$_\oplus$]                  & 1.67      & 2.61    \\
Mass   [M$_\oplus$]                  & 4.82      & 8.92     \\
Surface gravity  [m/s$^{2}$]         & 17.1      & 12.43    \\
Orbital Period [day]                & 11.0      & 32.94     \\
\hline
\end{tabular}
\label{tab:k2_values}
\end{table}

One orbital period ($P_{\rm orb}$), which is a year by convention, is
devided into 12 equal sized months.  For K2-18b, one year then
correspond to $\sim 32.94$\,Earth days, and a month to 2.74\,days.
Simulations were performed with a time step of 0.002 P$_{{\tt rot}}$,
and therefore varies between $\sim 0.0007$ months and $\sim 0.0036$
months.

\begin{figure}[h]
    {\includegraphics[scale=0.35]{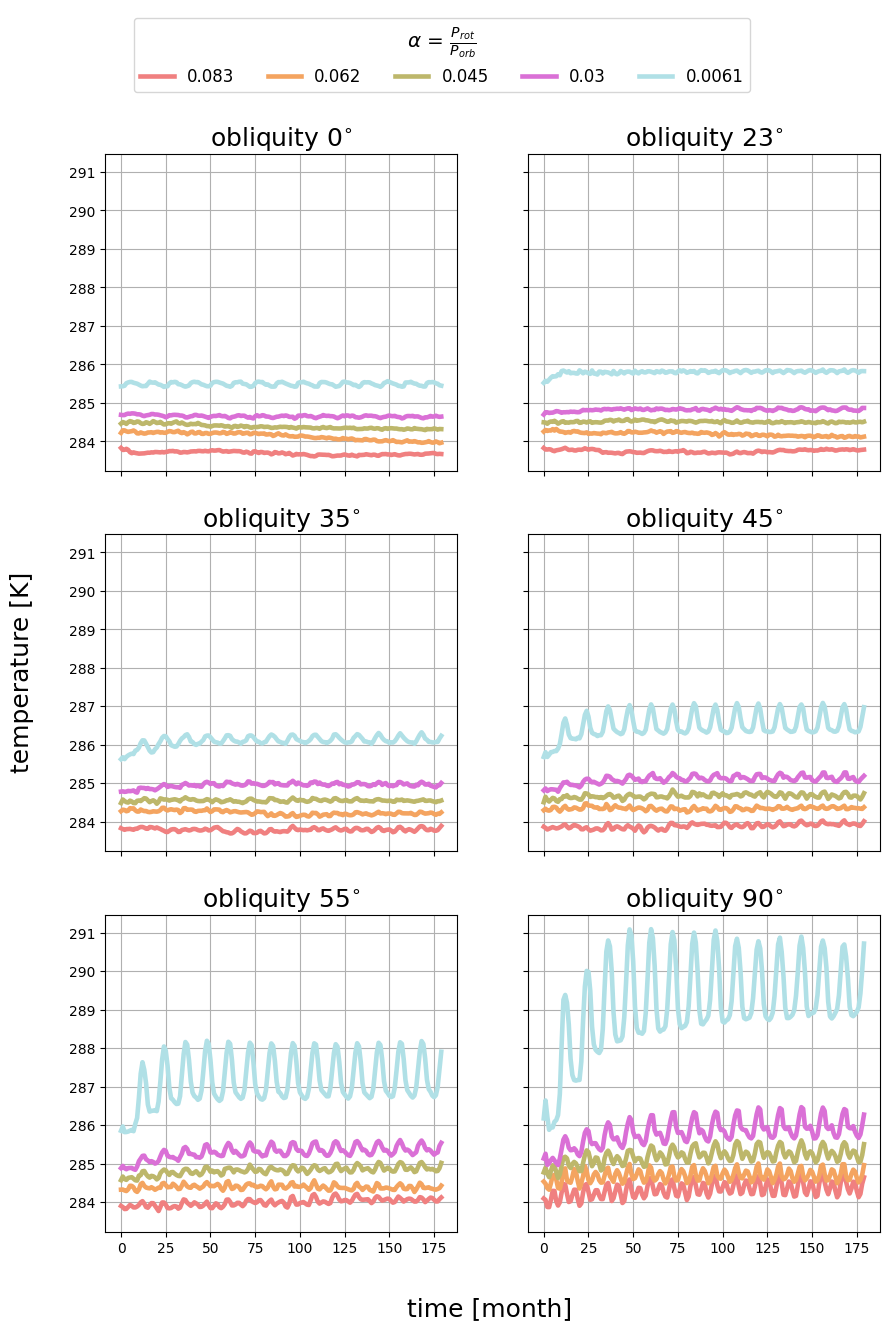}}
    \caption{Time evolution of the global average sea-level
      temperature for K2-18b over 180 months. Each frame represents a
      specific value of obliquity. Line colors represent different
      values of $\alpha$. The effect of its orbital eccentricity (of
      only $e=0.09$) is noticeable as the 12-month periodic variations
      in temperature. }
    \label{fig:temp_evolution}
\end{figure}

For most initial conditions, equilibrium in the atmosphere is reached
in 48 months (or 4P$_{{\tt orb}}$). But to ensure that all planet
atmospheres are relaxed under the effect of various rotation rates, we
run the code for 15 P$_{{\tt orb}}$ to achieve a stable equilibrium
atmosphere.  We subsequently instantaneously introduce the planet's
obliquity and run the simulations for another 15 P$_{{\tt orb}}$.  In
\cref{fig:temp_evolution} we present the last 15 P$_{{\tt orb}}$ of
the simulation in which the planet's rotation speed and obliquity are
set to the adopted values.

We introduce the parameter $\alpha \equiv P_{\rm rot}/P_{\rm orb}$,
representing the ratio between the rotational period
(P$_{{\tt rot}}$) and the orbital period (P$_{{\tt orb}}$) of the
planet. Note that for Earth $\alpha \simeq 0.0027$.  Higher values of
$\alpha$ lead to lower rotational speed for the planet (at the same
orbital period).  A value of $\alpha \to 1$ indicates synchronous
rotation, as in a tidally locked planet.

The atmosphere model became unstable for $\alpha > 0.083$ due to
extreme turbulence that drive high eddy kinetic energy.  We,
therefore, were unable to achieve lower rotation speeds, even after
reducing the time-step in the numerical solver by a factor of 30 to
0.00067 P$_{{\tt rot}}$.

We therefore limit our analysis to relatively low-values of
$\alpha$. This should not pose an interpretation problem for the wider
orbital separations in our sample, but limits the validity of our
calculations for tighter orbits, which are more likely to lead to
tidally locked planets.

Output is generated every $1/12$-th of an orbital period
(P$_{{\tt orb}}$), or monthly in terms of code units. For the
simplicity of post-processing, the outputs files are written per
orbital period, producing a total of 15 data files for each 15 year
simulation. The default output is a grid-based data format that can be
processed and analyzed using the \citep[\texttt{Grid Analysis and
    Display System (GRADS)},][]{GrADS} for the Earth surface.

\section{Results}\label{sec:results}

\subsection{Global climate for super Earth planets}\label{Sect:Results:WaterEarth}

We start by focusing on $8.92$\,M$_\oplus$ super Earth planets, later
in \cref{sec:discussion} we relax the planet mass and study the entire
range down to $1.0$\,M$_\oplus$.

In \cref{fig:temp_evolution} we present the time evolution of the
global averaged monthly mean temperature over the entire simulation
(15 P$_{{\tt orb}}$) for all values of the Obliquity and $\alpha$.
Each panel corresponds to a specific value of the obliquity, and each
curve represents one value of $\alpha$. Planets with a smaller value
of $\alpha$ (faster rotational velocity) have higher global averaged
temperatures. These trends are independent of obliquity.

For zero obliquity (aligned planet and orbital angular momentum axes),
all planets are able to maintain a stable global temperature. Lower
rotation velocity leads to a more effective cooling of the planet and
a lower average temperature. The difference in terminal average
temperature (after 180 months) is $\sim 2^\circ$.

For higher obliquities, in particular at 55$^{\circ}$ and
90$^{\circ}$, seasonal temperature variations become evident, with
planets having a lower $\alpha$ exhibiting more pronounced temperature
oscillations; more rapidly rotating planets have stronger seasonal
variations. This agrees with previous studies that shown atmospheric
heat transport is more efficient on slow rotating planets, and
increasing the rotating rate would lead to a greater amplitude of the
seasonal temperature variation, affecting the seasonal cycle, as the
formation of localized eddies are enhanced by stronger Coriolis
effects
\citep{2011Icar..212....1E,2015ApJ...804...60K,2022AGUA....300684G}

Planets with a 90$^{\circ}$ obliquity, have an average difference in
temperature of $\sim 7^{\circ}$ between the slowest and the most
rapidly rotating planets. Variations in temperature is seasonal but
also depends on the planet's rotation speed.

\begin{figure*}[h]
\centering
{\includegraphics[scale=0.32]{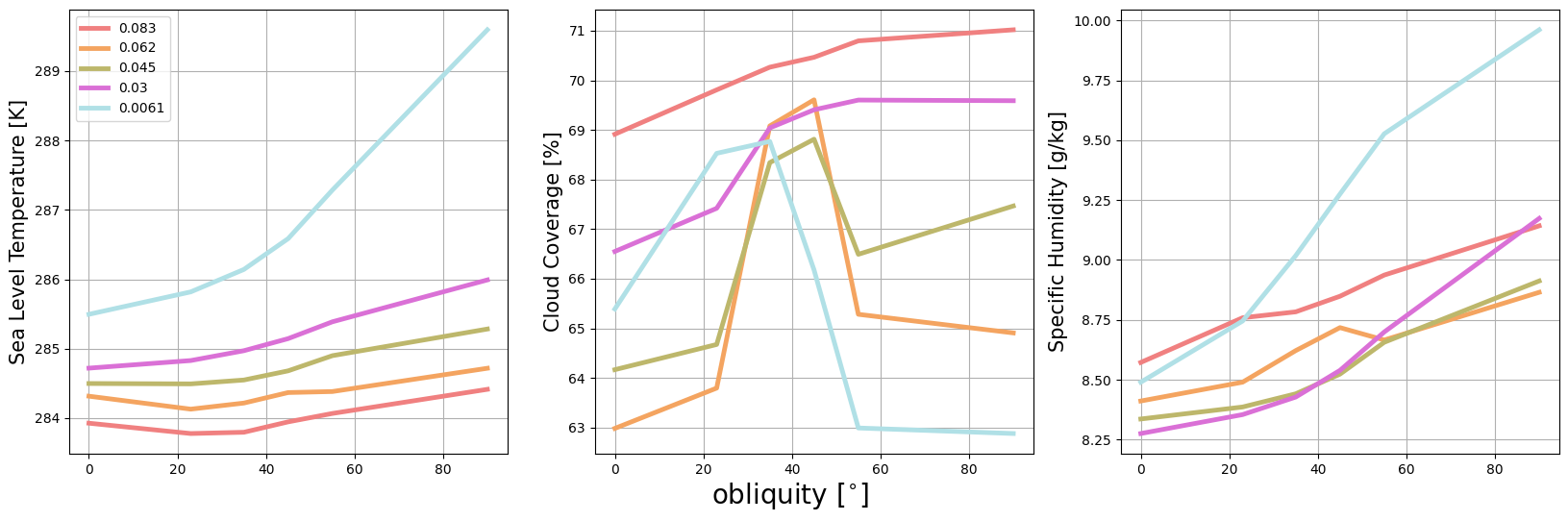}}
\caption{Global average values of the sea-level temperature (left
  panel), deep cloud coverage (middle), and specific humidity (right)
  as a function of obliquity and $\alpha$.  Each panel has several
  curves for each value of $\alpha$; the legend is in the left-most
  panel.  }
         \label{fig:global_surface}
\end{figure*}

In \cref{fig:global_surface}, we present the sea-level temperature,
deep cloud coverage, and specific humidity as a function of obliquity
and $\alpha$.  The measurements are time averages over the last two
orbital periods for the simulations with an equilibrium atmosphere.

Sea-level temperature increases with obliquity for all rotation
speeds. This rising temperature is more pronounced for planets that
rotate faster (lower $\alpha$).  In particular for $\alpha = 0.0061$
(cyan) the sea-level temperatures increases from $t=285.5^\circ$ for
an obliquity of 0$^{\circ}$ to $t=289.5^\circ$ for 90$^{\circ}$
obliquity.  For $\alpha = 0.083$ (red curve) the trend is similar, but
the temperature rises only by about 1$^{\circ}$.

The deep cloud coverage (middle panel in \cref{fig:global_surface}),
presents a more complex relationship. Here we consider deep cloud as
clouds with considerable vertical developments, including cumulonimbus
and tall cumulus clouds which can extend to the tropopause. An initial
increase in cloud coverage with increasing obliquity is found for all
rotational speeds. For all except for $\alpha=$0.083, a peak is
reached for an obliquity between 35$^{\circ}$ to 45$^{\circ}$.  This
trend breaks for $\alpha = 0.083$, for which the cloud coverage shows
a gradual increase with obliquity, but with a decreasing gradient for
high obliquity.

Surface humidity (right-most panel in \cref{fig:global_surface}) shows
a similar trend as temperature, illustrating how the specific humidity
follows the higher global temperatures by enhancing the atmosphere's
capacity of the warmer air to hold moisture.

\subsection{Zonal temperature variations} \label{sec:results-temp_var}

\begin{figure*}[h]
\centering
    {\includegraphics[scale=0.46]{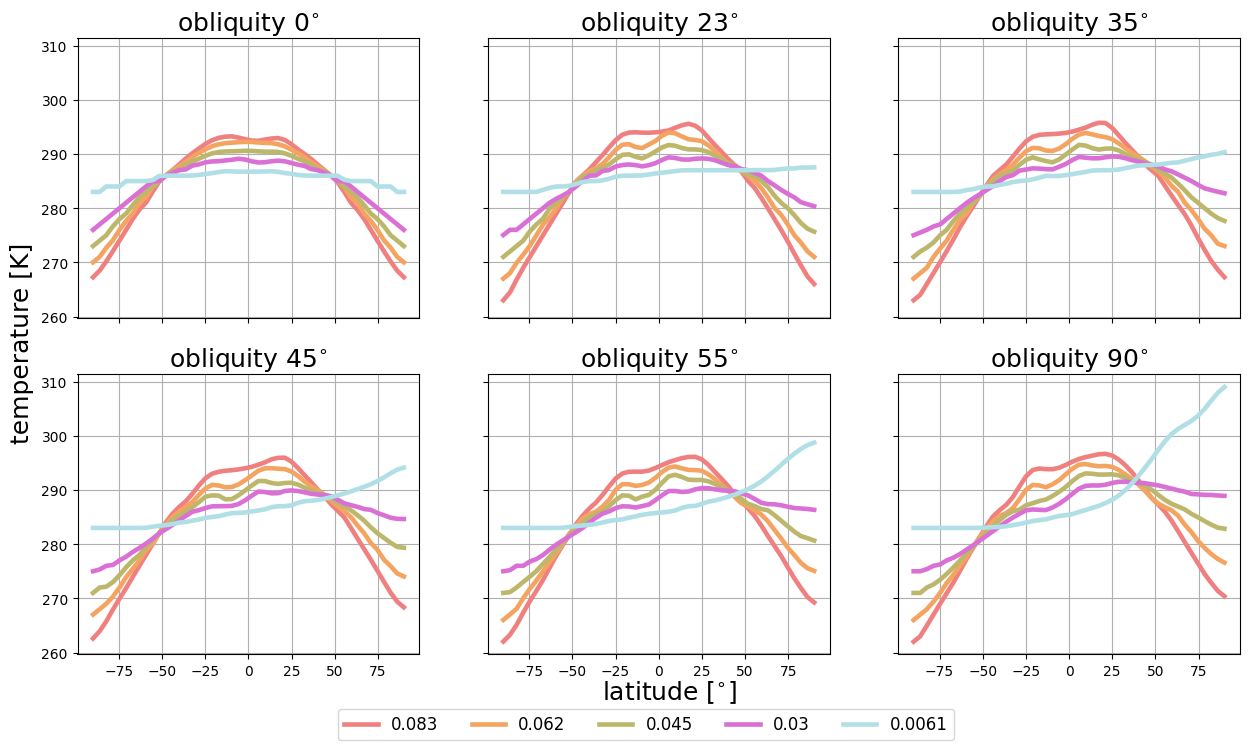}}
      \caption{Latitudinal profiles of sea-level temperature for
        different values of the obliquity and $\alpha$. Each panel
        represents a specific obliquity and the curves are colored
        according to $\alpha$ using the same color scheme for each
        panel. The legend is presented along the bottom of the figure.
        Negative latitudes are in the Southern hemisphere and positive
        values represents the Northern hemisphere; the equator is at
        0$^{\circ}$. The presented data gives the time average of the
        last two orbital periods in the calculation of the planet's
        atmosphere.}
         \label{fig:zonal_temp}
\end{figure*}

In \cref{fig:zonal_temp}, we present the longitudinally averaged
sea-level temperature as a function of obliquity and rotation
speeds. For this purpose, we divide the planet into 48 zones, from the
north to the south, with 3.75$^{\circ}$ per zone.  We calculate the
time-averaged mean temperature over each latitudinal zone.

The temperature profiles are symmetric around the equator for planets
for which the angular momentum axis is aligned with their orbital
angular momentum axis (zero obliquity), irrespective of $\alpha$.
Temperature is, as expected, highest at the equator, and decrease
towards the poles. The difference between polar and equatorial
temperature is largest for the slowest rotating planets.  We find
similar trends in the sea-level temperature and humidity.

\subsection{Polar temperature asymmetry}

\begin{figure}[h]
    \centering
    \includegraphics[scale=0.47]{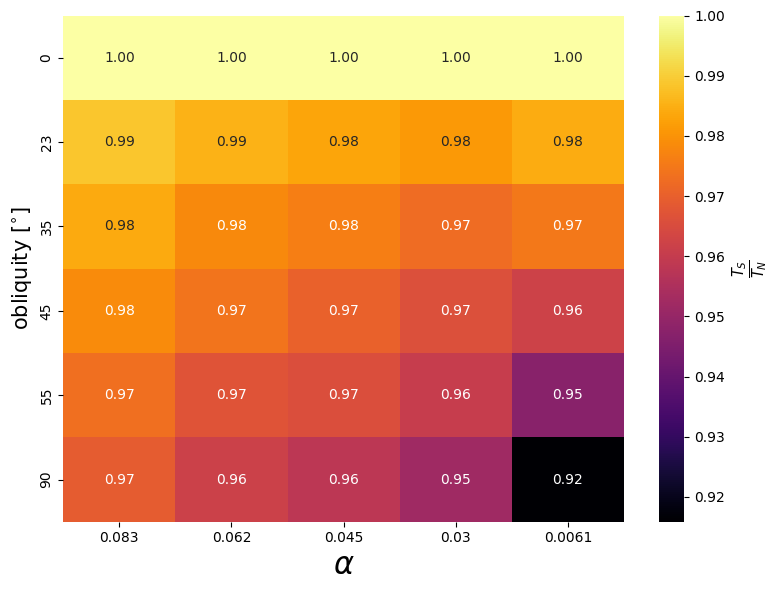}
    \caption{Heat map for the dependency of the polar temperature
      ratio $T_{\rm S}/T_{\rm N}$ as a function of obliquity and
      $\alpha$.  The temperatures were measured at at sea level ($P=
      1013$\,mbar). Colors indicate the temperature ratio, with a
      color bar on the right illustrating the range from 0.92
      (darkest) to 1.00 (lightest). Larger deviation from unity
      indicates greater asymmetry.}
    \label{fig:alpha_map}
\end{figure}

We further quantify the polar temperature asymmetry in
\cref{fig:alpha_map}, where we present the south-north
pole-temperature ratio, \TSTN. For consistency, these values are
averaged over the latitudinal zone from 86.25$^\circ$S to 90$^\circ$S
for the south-pole region, and from 86.25$^\circ$N to 90$^\circ$N for
the planet's north-pole region.  For planets with zero obliquity,
$\TSTN = 1$ across all values of $\alpha$, indicating a symmetric
temperature distribution between the poles.  The highest obliquity
($90^\circ$) shows the widest variation in \TSTN, ranging rom 0.97 for
slow rotating planets to 0.92, is consistent with the latitudinal
temperature profiles in presented in \cref{fig:zonal_temp}.

For an obliquity $\apgt 23^\circ$, an asymmetry develops in the
south-north pole sea-level temperature.  The effect becomes more
pronounced for higher obliquity and more rapidly rotating planets.
The more extreme polar temperature indicates an effective
redistribution of heat towards the poles, whereas heat flows away from
the equator for more rapidly rotating planets. For the oblique
planets, the higher temperature at the north pole compared to the
south pole is an interesting phenomenon. We tested the origin of this
temperature asymmetry for the fastest rotating planet with the highest
obliquity, for which the effect is most pronounced.

For larger obliquity, we notice an systematic difference in north-pole
versus south-pole temperature.  This trend becomes stronger for
smaller values of $\alpha$, when the planet is rotating faster. The
strongest trend is visible for $90^\circ$ obliquity for the lowest
value of $\alpha$ (see the aquamarine-colored curve at the
bottom-right panel in \cref{fig:zonal_temp}).  The most rapidly
rotating planets have no peak temperature at the equator (except for
aligned planets). Even a slight reduction in the planet's rotation
speed breaks this trend. This systematic north-south pole temperature
difference is probably also the cause of high lateral turbulence and
extreme eddy kinetic energies in the atmosphere and ocean, which
prevent us from performing simulations with even higher rotation
rates. The maximum north-south pole temperature difference exceeds
25\,K. With extreme high polar temperatures in excess of 310\,K. Local
conditions may drop below the freezing point of water, or exceed its
boiling point. The wide range in average temperature is therefore
probably limited to around 35\,K. Such a steep temperature gradient
between pole and equator is reached for 90$^\circ$ obliquity.

To further explore this phenomenon, we perform an extra series of
calculations. We simulated the planets with identical obliquities but
towards the opposite direction by starting the simulations with an
offsets of a quarter and half a orbital period, and turning the planet
around. But the asymmetry persists; when flipping the planet, the
north-south asymmetry persists reversed (see
also\,\cref{sec:discussion}).  The asymmetry in south-north polar
temperature appears to be related to small initial variations in the
polar temperature. For the simulations where one of the poles is
directed towards the star, the initial directed hemisphere heats up at
the start of the simulation, but remains slightly hotter than the
other pole after one complete revolution.  This asymmetry persists
throughout the simulation, and increases until the steady-state
atmospheric conditions are reached (for the planets with the highest
obliquity, this happens already after $\sim 72$ months, or 6 initial
orbits).

In the simulations where the rotation axis of the planet started
pointing along the planet's orbit, the asymmetry is initiated by the
first polar passage.  The temperature differences is introduced by a
small but finite temperature difference between south and north
pole. This slight temperature difference is introduced in the first
quarter orbit, and is magnified until multiple equilibria are reached,
and persists afterwards.

\begin{figure}[h]
    \centering
    \includegraphics[scale=0.47]{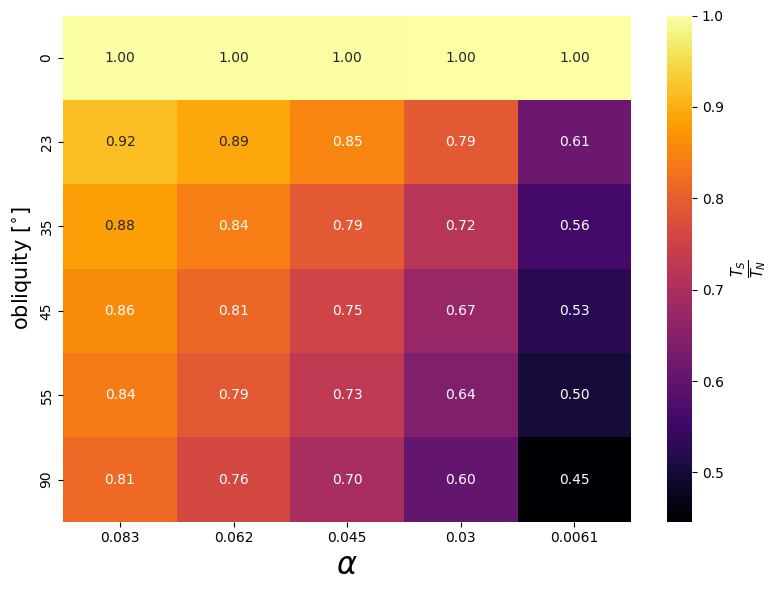}
    \caption{The dependency of the polar temperature ratio \TSTN\, as
      a function of obliquity and $\alpha$ in the base of the
      stratosphere at a pressure of $P = 30$\,mbar.}
    \label{fig:alpha_strat}
\end{figure}

In \cref{fig:alpha_strat} we present a similar analysis in terms of
\TSTN\, but at the base of the stratosphere.  High up in the planet's
atmosphere the temperature asymmetry, already discussed at sea-level
in \cref{fig:alpha_map}, is even more pronounced.  At high altitudes,
the North pole can become twice as hot as the South pole.

The lower atmosphere of an aqua-planet is more strongly affected by
the surface water than higher in the atmosphere.  The surface water's
high specific heat capacity regulates the redistribution of heat.
This is a direct result of the higher efficiency of heat transport in
the ocean than in the atmosphere.  The greater atmospheric humidity at
lower altitudes also facilitates heat distribution through greenhouse
effects and the enhanced irradiation absorption. The profound
asymmetry at higher altitudes can then be attributed to the lack of
these regulating mechanisms, resulting in more pronounced temperature
differences between the poles.

\subsection{Cloud coverage}

\begin{figure*}[h]
    \centering \includegraphics[scale=0.28]{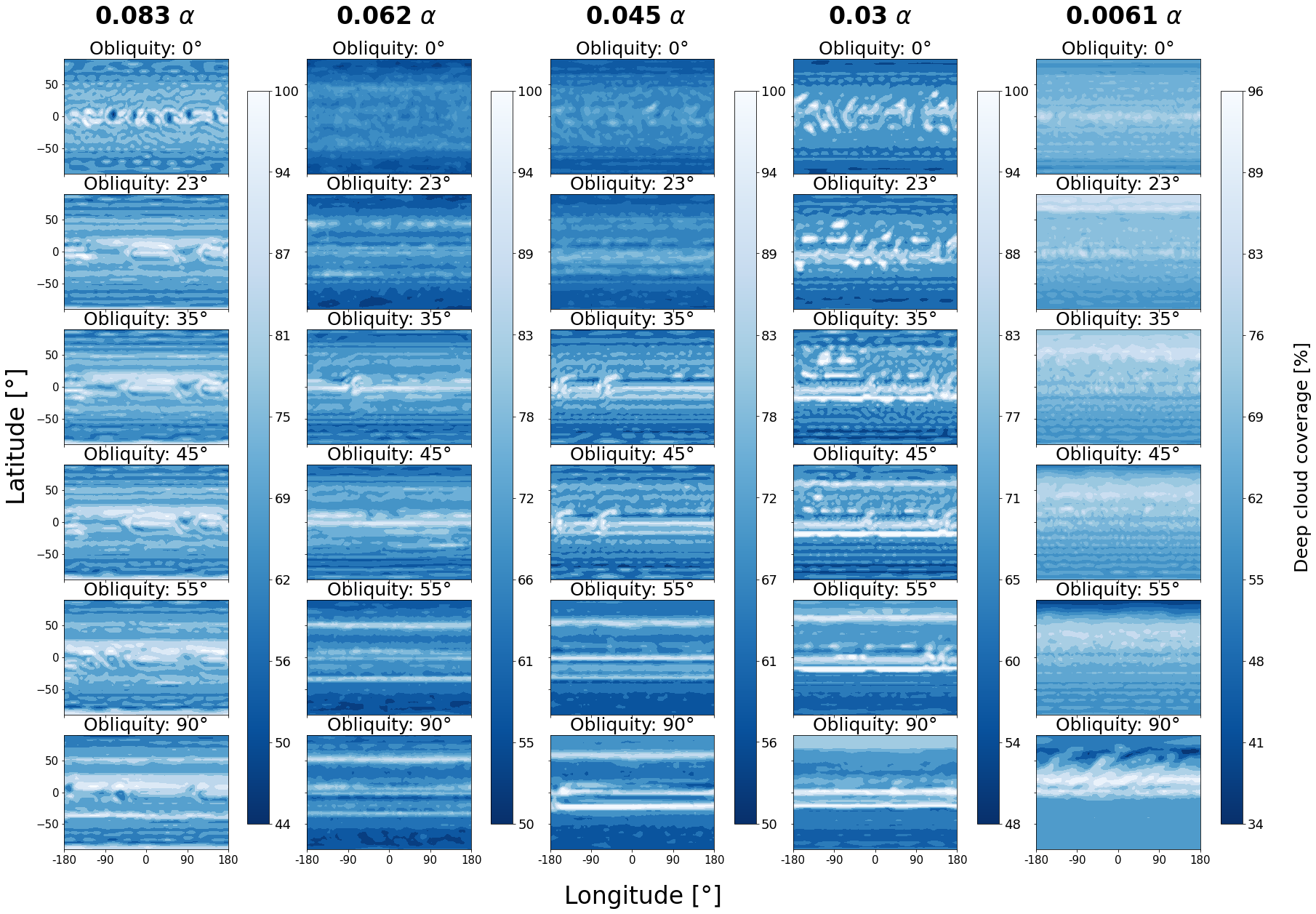}
    \caption{Maps of deep cloud coverage for each $\alpha$ simulation
      for all obliquity cases. Each column has independent color
      bars. Distinct longitudinal cloud bands appear at higher
      obliquity with an overall greater cloud coverage. }
    \label{fig:cloud_all}
\end{figure*}

To further understand the polar-temperature asymmetry, we present in
\cref{fig:cloud_all} the cloud coverage for each of the
simulations. We observe that for all values of $\alpha$ the $0^\circ$
obliquity planets show a symmetric cloud coverage across latitude and
a constant behavior along longitude.

When increasing obliquity for the slowest rotating planets
($\alpha=0.083$) we observe a change in equatorial turbulence and
a thinning of the cloud cover over both polar regions.  When the
planets' rotation speed increases the equatorial cloud cover
increases, and becomes laminar. The cloud coverage at the south pole
decreases for faster rotating planets, whereas the north pole
preserves a thick cloud coverage. The planets with the highest
rotation rates exhibit different trends with increasing obliquity,
where we observe the development of a laminar flow that breaks at the
north pole once the obliquity exceeds $\sim 45^\circ$.

We do not see evidence for a hexagon polar configuration, as observed
at Saturn's poles \citep{1988Icar...76..335G}.  In our relatively
low-resolution simulations, such signatures would be hard to identify.
We do notice some regular cloud patters along latitude $\sim 84^\circ$
for both poles in the low-obliquity models.  This could indicate the
presence of circumpolar cyclones, as observed around Jupiter's poles
\citep{2018Natur.555..216A}.  This phenomenon is most clearly visible
at the $\alpha=0.083$ panel in \cref{fig:cloud_all}. Such regular
pattern could be the result of vorticity dynamics, as is the case for
Jupiter \citep{2021NatGe..14..559G}.

\subsection{Mass dependency of the temperature asymmetry}

\begin{figure}
    \includegraphics[scale=0.4]{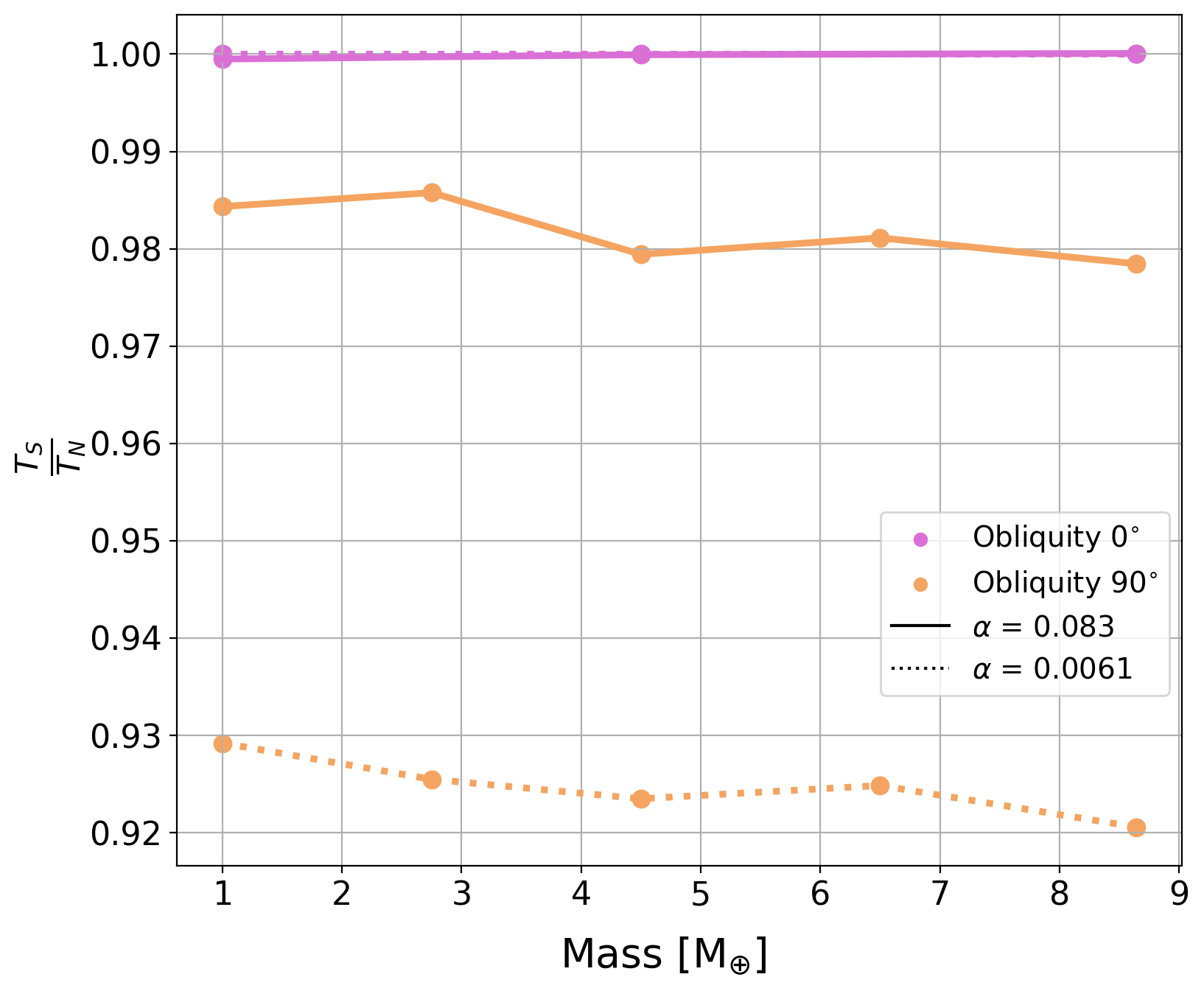}
    \caption{South-north-pole temperature ratio as a function of
      planet mass.  The bullet points indicate the result of the
      performed calculation, and the lines are drawn to guide the
      eyes.  solid lines are drawn for values of $\alpha=0.083$, and
      dotted lines for $\alpha=0.0061$. For $0^\circ$ obliquity the
      dotted and solid covers are indistinguishable.  }
    \label{fig:mass_plot}
\end{figure}

The calculations so far were performed with a specific choice of the
planet mass, that of K2-18b.  To further explore the generality of the
results on cloud coverage and temperature asymmetry, we vary the
planet mass, but keep the other parameters, including the density of the planets, constant. Again, we cover the entire parameter range in obliquity and $\alpha$, analyzing the south-north-pole temperature ratio.  The results are presented for the least and most rapidly rotating planets, and for the obliquities of $0^\circ$ and $90^\circ$. The results of these calculations are presented in \cref{fig:mass_plot}.

The planets with $0^\circ$ obliquity show no variation in the polar
temperature irrespective of mass or rotation speed.  The two curves in
\cref{fig:mass_plot} are indistinguishable.  For planets with
$90^\circ$ obliquity, we notice a decreasing trend in \TSTN\, with
mass, but the most profound difference is found between slowly and
rapidly rotating planets, which are offset by a considerable
margin. The earlier observed trend in the asymmetry of the temperature
at the poles for highly oblique planets persists with mass.

\section{Discussion}\label{sec:discussion}

In our adopted low eccentricity of $e=0.09$, the star's irradiation is
varies by about 44 per cent. The planet rotation then prevents the
lateral transport of heat across the planet atmosphere, which, as a
consequence, becomes laminar with a wide turbulent region around the
equator and the poles.  Such turbulence gives rise to a broad region
of almost constant temperature between a latitude of $-30^\circ$ and
$+30^\circ$ with varying cloud coverage.

Once the planet's obliquity increases a persistent temperature
asymmetry arises; one of the poles becomes hotter than the
other. This asymmetry is enhanced by an asymmetry in polar cloud
coverage, and persists across planet mass.  The temperature asymmetry
is supported and maintained by the laminal atmospheric conditions,
strengthened by the planet's rotation speed.

Our findings are supported by earlier simulation for a water Earth by
\cite{Ferreira2014} and \cite{Lobo2020}, who find annual shifts in the
averaged stellar radiation towards the poles for planetary obliquity
greater than 54$^{\circ}$. This tendency for polar heating in
high-obliquity planets was also found in our validation runs with
water Earth models in \cref{Sect:Results:WaterEarth}. However, we did
not find evidences for polar heating in the zonal temperature profiles
for our equivalent planet in \cref{fig:zonal_temp}.  Instead, we find
a profound asymmetry in temperature and cloud coverage between the two
hemispheres that increases with obliquity.  The northern hemisphere,
in particular near the pole, warms up compared to the equator, whereas
the southern hemisphere cools down. The pole that heats up depends on
the variations in the model.  Verification runs with reversed
rotational axes, yielding an obliquity of 360$^{\circ} - o$ (where $o$
represents the originally tested obliquity of 0$^\circ$, 23$^\circ$,
35$^\circ$, 45$^\circ$, 55$^\circ$, 90$^\circ$), produce a hotter
southern hemisphere with the same degree of asymmetry.

The temperature asymmetry persists across a pressure range from 1013
to 30 mbar.  When comparing the results in \cref{fig:alpha_map} and
\cref{fig:alpha_strat}, we observe an increased range in the
temperature asymmetry in the higher atmosphere.  For an aquaplanet
with physical parameters similar to K2-18b but with an Earth-like
atmosphere, the effects of high obliquity is less profound at slower
rotation rate ($\alpha \ge 0.03 $), and in this case a global
equilibrium climate can be achieved.

In earlier work using a climate model of intermediate complexity, with
a similar imflux of 1365\,W\, m$^{-2}$.  \cite{2014EGUGA..1615068L}
and \cite{2015P&SS..105...43L,Kang2019} find a similar trend in the
lateral surface temperature \citep[see figure 3
  in][]{2015P&SS..105...43L}, but they do not find a temperature
asymmetry in highly oblique ($\apgt 80^\circ$) Earth-like
planets. Although the resolution of their grid (on the order of
$64\times32$ with 10 lateral layers in \cite{2015P&SS..105...43L} and
one layer in \cite{Kang2019}) is comparable to our calculations.  The
planets in \cite{2015P&SS..105...43L} are rapid rotators, whereas ours
includes slow rotators; the length of the day is 24 hour in
\cite{2015P&SS..105...43L} whereas ours varies from 4.8 to 66 hours.
\cite{Vallis:2009} point out that different rotation rates may have a
large impact on the meridional heat transport.  Also
\cite{2022ApJ...940...87K} find no appreciable asymmetry between
northern and southern hemisphere temperatures for their rapidly
(synchroneously) rotating Earth-like planets.  Like our model, they
adopted a vertical slab ocean but without lateral energy transport.
Another study shows a clear north-south asymmetry in precipitation
\citep{https://doi.org/10.1029/2018MS001313} which they attribute to
the structure of the overturning cell and associated shift in the
annual mean intertropical convergence zone.  Note that a slight
difference in precipitation and wind speed for simulation of a slowly
rotating K2~18b-like planet was observed in \citep[][see their
  fig.~10]{2021A&A...646A.171C}.

The systematic temperature difference between the two poles can result
from separate circulation cells which prevent thermal energy exchange
across the equator and preserve the thermal perturbations in our high
obliquity simulations. These Hadley cells cause the observed asymmetry
across multiple climatic observables.  As a result, different
equilibria can exist simultaneously in the northern and southern
hemispheres. This process could be effective in slowly rotating
planets, but absent in rapid rotators, explaining why a similar
asymmetry was not found in \cite{2014EGUGA..1615068L} and
\cite{2015P&SS..105...43L,Kang2019}.  Our model has a reflection
symmetry $f(-x, t) = -f(x,t)$ with respect to the equator probably
resulting in a pitchfork bifurcation, which leads to different
equilibrium solutions in the two hemispheres.  Such multiple
equilibria in a climate system can result from bifurcations induced by
non-linear processes in the model \citep{multi_equi}. In the case of
\texttt{ExoSPEEDY}, such non-linearities are introduced in the
computation of the radiation and surface fluxes of the atmospheric
model.

Recently \cite{hammond2024} used the 3D GCM model ExoCAM1
\citep{exocam} to analyze the climate of TRAPPISI-1e for a range of
obliquities for a tidally locked planet. They also found zonal
temperature variation with increasing obliquity and decreasing
atmospheric pressure.  In our case, the effect also becomes more
pronounced at higher altitudes where the pressure is lower. The
thermal emission difference between the two poles is then easiest
detected at the top of the atmosphere, rather than at the bottom.  A
temperature difference of $>25^\circ$\,C could be detectable across
the planet's surface. In that case, it could be possible to develop a
method to infer the planet's obliquity from such a temperature
asymmetry. Since cloud coverage correlates with temperature, an easier
way to delineate the planet's obliquity may be by observing the
planet's albedo difference in the two hemispheres \citep[see
  also][]{2021A&A...646A.171C}. A difficulty here is, that one may not
be able to identify the hemispheres before doing such a study, and the
planet's equator becomes part of the solution.

It is unclear what would happen if we introduce land-mass or a
biosphere on the planet surfaces.  Cloud formation in such a case, is
could be quite different because many of the nuclei for cloud
formation on Earth have a biogenic origin.

\subsection{Bifurcation of atmospheric circulation}\label{sec:dis_bifurcation}

Our findings of a persistent north-south asymmetry in the polar
climate of oblique, rapidly rotating aqua-planets align with the
concept of atmospheric bistability, which are regions with multiple
solutions, studied by earlier efforts focusing on the atmospheric
circulations of terrestrial planets \citep{Edson2011, Herbert2020}. It
was pointed out by \citep{Edson2011}that the circulation patterns for
tidally locked planets depend on the rotation period. For aquaplanet,
the transition between different circulation regimes and multiple
equilibria can occur between a 3-day and 4-day period. More recently,
\cite{Sergeev2022} show that the atmosphere of TRAPPIST-1e can stably
exist in two dynamically distinct states, either dominated by an
equatorial jet or by mid-latitude jets, despite identical forcing and
boundary conditions, yielding different temperature, wind, and cloud
distribution for the two regimes. The bifurcation arises from early
divergences in diabatic forcing, amplified by nonlinear dynamical
feedbacks such as wave–mean flow interactions and differential cloud
radiative effects, and is sensitive to small perturbations in the
initial conditions. Similarly, our simulations show that hemispheric
asymmetries can be self-reinforcing and lead to a divergent, yet
stable climate state. These parallels suggest that the bifurcation
observed in our model is not a numerical artifact. By imposing a high
obliquity as the initial condition, we effectively assigned the two
hemisphere with different level of stellar irradiation and different
initial temperatures, leading to later divergences. Our results
complement this picture by demonstrating that rapid rotation and high
obliquity on non-tidally locked water planets can also establish
atmospheric multi-equilibria by driving a persistent north–south polar
temperature asymmetry.

\subsection{Applications and limitations} \label{sec:dis_app}

Observing a planet's obliquity is difficult.  Observations of
exoplanets have primarily focused on measuring stellar obliquity and
other, more accessible, orbital parameters, such as eccentricity,
orbital period, mass, and radius.  Measuring surface temperature
profiles through thermal emissions is well-developed and widely
practiced \citep{webb_temp}.

The range over which we varied $\alpha$ is limited to numerical
constraints, rather than physical ones. Improving this, however, may
not be trivial as the reason for the code's inability to resolve the
fastest rotating planets may not be solved by increasing the spatial
or temporal resolution of the model.  Using the Community Atmosphere
Model \citep{Neale2010CAM4}, \cite{2020A&A...643A..37Y} managed to
simulate tidally locked planets.  They demonstrate that massive storm
complexes may arise in zero-obliquity tidally-locked Earth-like
exoplanets in the habitable zone at a similar insolation range of
$I_{irr} = 1300$ to $1800$\,W m$^{-2}$.

The problem may arise from
missing transport processes and unresolved turbulence.  The eddy
kinetic-energy threshold for \texttt{ExoSPEEDY} is exceeded for
$\alpha \ge 0.1$. Using the solar system as a reference, the
rotational period for the giant planets ranged from approximately 10
to 20 hours, resulting in an order of magnitude smaller $\alpha$
values, from $\sim 10^{-4}$ for Jupiter to $\sim 10^{-5}$ for Neptune.
Extrapolating our results to the solar system's giants is not trivial
because their larger distance to the Sun makes these planets dominated
by ice rather than liquids.

\subsection{Temperature asymmetry in Uranus} \label{sec:uranus}

Although high obliquities seem exotic, in the Solar system two
planets, Venus ($177.4^\circ$) and Uranus ($97.86^\circ$), have high
obliquity.  Possible even Earth had a high obliquity in the past
\citep{2016Natur.539..402C}.  The obliquity of a planet may be rather
fundamental to its formation process \citep{2006Natur.440.1163B}, and
having an indirect diagnostic to find exo-planet obliquities could be
quite effective in constraining their formation histories and early
evolution.

The only spatially resolved planet where we could expect to observe an
asymmetry in the temperature in the northern and southern hemisphere
is Uranus. With an average temperature of about $78^\circ$K, Uranus is
the coldest planet in the Solar system. It exhibits enormous
temperature variations along its orbit, similar to the yearly
temperature variations ($\apgt 2^\circ$) we see in the highly oblique
planets in \cref{fig:temp_evolution}.

In spatially resolved thermal emission from the planet Uranus between
2003 and 2011 \cite{ORTON201594} found a hemispheric temperature
difference between the north and the south.  Their results are
consistent with the 1986 Voyager IRIS experiment.  The data was taken
around the time Uranus passed its equinox in 2007, but acquiring an
orbit-average temperature of Uranus would take about 84 years.

The data by Voyager and \cite{ORTON201594} was taken in the upper
troposphere of Uranus, at an atmospheric pressure of around the 70–400
mbar. This is a bit lower in the atmosphere as the calculations we
present in \cref{fig:alpha_strat}.  For 90$^\circ$ obliquity at
30\,mbar we find a temperature difference of \TSTN\,$\sim 0.81$ and
$0.45$.

At planetographic latitude $60^\circ$S, and $30^\circ$N,
\cite{ORTON201594} measure a temperature of about 51\,K, whereas at
$30^\circ$S, $60^\circ$N, they measure a temperature of $49$\,K and
$46$\,K, at a wavelength of 18.7 $\mu m$ and 22.0 $\mu m$,
respectively.  This leads to a relative temperature difference of
$\TSTN \sim 0.86$ to $0.96$. Although hard to compare directly with
our simulations, Uranus also seem to exhibit a latitudinal temperature
asymmetry.

\subsection{The exo-planets K2-18b and TOI-1452b} \label{sec:K2-18b}

For hycean planets with a water mass fraction of 10-90\%, the boundary
between the hydrogen/helium (H/He) envelope and the H$_2$O layer must
exist at pressures and temperatures that support the liquid phase of
water. This H/He-H$_2$O boundary (HHB),
defines the surface of the planet. Irrespective of the H-rich
atmosphere, these planets are the closest example of an aqua world
today.

Of particular interest in this family of aqua planets is K2-18b
\citep{k2_discovery}, with an irradiation of $I_{\star} = 1368
^{+114}_{-107}$ W m$^{-2}$ and a mass of $\sim 8.92$ M$_{\oplus}$
\citep{k2_discovery}.  Although, recent transmission spectra, observed
with JWST NIRISS and NIRSPEC, have revealed this atmosphere to be rich
in methane (CH$_4$) and carbon dioxide (CO$_2$) (with non detection of
ammonia (NH$_3$)), the underlying world is suggested to host a liquid
water-rich ocean \citep{Madhusudhan_hycean}.  K2-18b is therefore
classified as a liquid H$_2$O-covered or Hycean sub-Neptune with a
H-rich atmosphere \citep[][rather than $N$-rich in our
  model]{fulton_2017,zeng_2019,otegi_2020}.

Internal modeling of K2-18b using the \texttt{HyRIS} model was
conducted by \cite{madhusudhan_k2ocean} to determine the depth of
potential liquid oceans on various sub-Neptunes, including K2-18b.
Based on the criteria for life on Earth, they assumed a surface
temperature T$_{{\tt HHB}}$ in the range of 273-400 K and with the
pressure P$_{{\tt HHB}}$ between 1 bar to 1000 bar for a habitable
Hycean planet. An ocean depth range of 140-180 km was found using the
pressure-temperature profile generated by the \texttt{GENESIS} model
from \cite{madhusuhdan2023a}, considering a T$_{{\tt HHB}}$ of
340K. When using an adiabatic profile with an ambient temperature of
250 K and a radiative-convective boundary at 100 bar, a shallower
ocean depth range of approximately 50–180 km was identified through
their model. An ocean depth range from 50-350 km was found across all
the pressure-temperature profiles, greater than Earth's average ocean
depth of 3.7 km \citep{NOAAocean}.
  
The planet TOI-1452b, is also of interest to our study since it is
considered habitable according to some criteria
\citep{2019ApJ...887L..14B,2023ApJ...948L..26H}.  It is a bit warmer
and about half the mass of K2-18b. Both planets are claimed to be
covered by a liquid ocean: for K2-18b this is probably a Hycean ocean,
whereas TOI-1452b may be water dominated.

Previous studies have assumed K2-18b to be tidal synchronized with
zero obliquity due to its proximity to the host star at 0.16 au
\citep{benneke2019b, Blain2021, madhusudhan_k2ocean, madhusuhdan2023a,
  Madhusudhan_hycean, Shorttle2024}. TOI-1452b is close to its parent
star, and could well be tidally locked too.

However, the discovery of K2-18c, with an inner orbit at 0.06 au and a
mass of 7.51 M$_{\oplus}$ \citep{cloutier2017}, adds complexity to the
K2-18 system. The presence of another mass in the system causes a
rotating planet to experience gravitational torque from both the host
star and the additional mass. The planet's spin axis precesses around
its orbital axis due to the host star's tidal influence, while the
orbital axis itself precesses around another axis under the influence
of the secondary mass. When the precession frequencies of the two axes
are comparable, a resonance can increase the planetary obliquity to
large values. This theory, known as the ``Colombo's Top'' model
\citep{colombo}, has been used to explain the obliquities of solar
system giants \citep{vokr2015, saillenfest2020, saillenfest2021,
  rogoszinski2020}.

The presence of an inner planet with a comparable mass could
potentially lead to a precession resonance for K2-18b, resulting in
non-zero obliquity. In the case of TOI-1452b, as part of a binary star
system, it would be possible for the secondary star to induce torque
on the planet and excite non-zero obliquities. Additionally, K2-18b's
non-zero eccentricity(e$\approx$ 0.09) makes it a candidate for
spin-orbit resonance, similar to Mercury. This could result in a
planetary rotation period different from its orbital period,
facilitating faster heat redistribution and producing distinct
atmosphere dynamics compared to a tidally-locked scenario.

The fundamental parameters for both planets are summarized in
\cref{tab:k2_values}. Although K2-18b and TOI-1452b are both
aquaplanet candidates with potentially non-zero obliquity, extending
our findings to these specific planets is not trivial as our parameter
space is still quite limited. The mass and orbital separation for
K2-18b and TOI-1452b could be partially synchronized, in which case
they both are spinning much slower than our adopted range.  In
addition, we adopted an Earth-like atmosphere dominated by nitrogen,
which is different from the H/He envelope on Hycean planets. Ice
formation was not considered in the ocean model, even though the
temperature at the south pole approaches $260$\,K, beyond the water
freezing point at 1 bar.

\section{Summary and Conclusion}\label{sec:conclusion}

We have studied the effect of planet mass, rotation and obliquity on
the climate of ocean worlds. These calculations were performed using
the global circulation model of intermediate complexity
\texttt{ExoSPEEDY}, with orbital separation of $\sim 0.16$ au and eccentricity $e=0.09$.

Our calculations start by reproducing the current Earth, and an
Earth-equivalent without continents, covered with water.  The model we
adopted compared favorably to earlier Earth-like models.  In a second
series of calculations we vary the planet obliquity from $0^\circ$ to
$90^\circ$, the planet rotation from $\alpha = 0.0061$ to $0.083$, and
the mass from 1\,M$_\oplus$ to 8.9\,M$_\oplus$.

We find that varying planet mass has only a minor effect on the
equilibrium atmospheric conditions of a rotating planet, but obliquity
and rotation period affect seasonal variations, mainly by inducing
systematic seasonal asymmetries in surface temperature and cloud
coverage.  These asymmetries influence short-term weather patterns as
well as long-term climate variability.

In general, a lower obliquity leads to more uniform stellar
irradiation across the planet's surface, leading to milder seasonal
variations, whereas these variations become more pronounced for higher
obliquity.  Previous simulations focused on Earth-like conditions
found similar trends that highly oblique planets experience more
severe seasonal variation \citep{Ferreira2014,Lobo2020}.

For highly oblique planets a systematic asymmetry in the average
temperature develops between the northern and southern
hemisphere. This temperature difference is largest when comparing the
polar temperatures, but persists across the entire hemisphere. The
asymmetry is strongest for fully tilted and rapidly spinning
planets. Slower spinning planets still exhibit the asymmetry, but less
pronounced.  The polar temperature difference diminished for lower
obliquity, to completely disappear for aligned planets.

The planet's cloud cover shows a similar trend, in terms of more
clouds around the poles (compared to the equator) for planets with
high obliquity. The difference in polar cloud coverage and
temperatures are probably connected, in terms that a thicker cloud
coverage leads to higher local temperatures.

The temperature asymmetry persisted across a pressure range from 1013
to 30 mbar, with the most extreme differences found at higher
altitudes of oblique fast rotators. At 30 mbar, the polar temperature
ratio \TSTN\, for a 90$^{\circ}$ obliquity planet with $\alpha =
0.0061$ reached as low as \TSTN$ = 0.45$; one pole more than twice as
hot as the other pole.  The polar temperature asymmetry is more
pronounced in the upper atmosphere.  At high atmospheric pressure,
near the planet liquid surface, the large heat-capacity of the ocean
and its efficient heat transport couples to the atmosphere, moderating
the temperature asymmetry. In the upper atmosphere this effect is much
weaker, leading to differences of the polar temperatures.

The latitudinal temperature profiles (Fig.\ref{fig:zonal_temp}) and
the longitudinal asymmetry \TSTN\, (Fig.\ref{fig:alpha_map} and
\ref{fig:alpha_strat}) confirm that increased obliquity and faster
rotation rates cause temperature peaks to shift towards one of the
poles.  The resulting hemispherical temperature asymmetry leads to a
comparable asymmetry in cloud coverage, which affects the planet's
albedo. A temperature gradient can then be observable as gradient in
the planet's albedo.  Detecting such a gradient, either in temperature
or in albedo, may provide an independent proxy for planet rotation
speeds and obliquity. These latter parameters inform us about
the tidal evolution of the observed planet.

\section*{Acknowledgements}
ChatGPT was used to help with \texttt{Python} and \texttt{FORTRAN}
debugging and corrections of grammar mistakes.  Special thanks to
Daphne Stam for correspondence and assistance with the
\texttt{ExoSPEEDY} code package, which played a crucial role in the
execution of simulations and data analysis.
In this work we used planetary data from \url{https://openexoplanetcatalogue.com/}.







\end{document}